# Nematicity in stripe ordered cuprates probed via resonant x-ray scattering**

A. J. Achkar, [1] M. Zwiebler, [2] Christopher McMahon, [1] F. He, [3] R. Sutarto, [3] Isaiah Djianto, [1] Zhihao Hao, [1] M. J. P. Gingras, [1,4,5] M. Hücker, [6] G. D. Gu, [6] A. Revcolevschi, [7] H. Zhang, [8] Y.-J. Kim, [8] J. Geck, [2,9] and D. G. Hawthorn [1,4*]

[1]Department of Physics and Astronomy, University of Waterloo, Waterloo, N2L 3G1, Canada

[2]Leibniz Institute for Solid State and Materials Research IFW Dresden, Helmholtzstrasse 20, 01069 Dresden, Germany

[3]Canadian Light Source, Saskatoon, Saskatchewan, S7N 2V3, Canada

[4]Canadian Institute for Advanced Research, Toronto, Ontario M5G 1Z8, Canada

[5]Perimeter Institute for Theoretical Physics, 31 Caroline St. N., Waterloo, Ontario, N2L 2Y5, Canada

[6]Condensed Matter Physics and Materials Science Department, Brookhaven National Laboratory, Upton, NY 11973, USA

[7]Synthèse, Propriétés et Modélisation des Matériaux (SP2M), UMR 8182, Université Paris-Sud, 91405 Orsay Cedex, France

[8]Department of Physics, University of Toronto, Toronto, M5S 1A7, Canada

[9]Paris Lodron University Salzburg, Chemistry and Physics of Materials, Hellbrunner Strasse 34, 5020 Salzburg, Austria.

*To whom correspondence should be addressed; E-mail: dhawthor@uwaterloo.ca.

**In underdoped cuprate superconductors, a rich competition occurs between superconductivity and charge density wave (CDW) order. Whether rotational symmetry breaking (nematicity) occurs intrinsically and generically or as a consequence of other orders is under debate. Here we employ resonant x-ray scattering in stripe-ordered (La,M) $_2$CuO $_4$ to probe the relationship between electronic nematicity of the Cu 3$d$ orbitals, structure of the (La,M) $_2$O $_2$ layers, and CDW order. We find distinct temperature dependences for the structure of the (La,M) $_2$O $_2$ layers and the electronic nematicity of the CuO $_2$ planes, with only the latter being enhanced by the onset of CDW order. These results identify electronic**

**nematicity as an order parameter that is distinct from a purely structural order parameter in underdoped striped cuprates.**

Key challenges in resolving the cuprate phase diagram are to identify the broken symmetries that generically occur and understand how they are interrelated. Recently, a series of measurements have shown translational symmetry breaking in the form of charge-density wave (CDW) order to be generic to underdoped cuprates (*1, 2, 3, 4, 5, 6, 7, 8, 9*). It has also been proposed that electronic nematicity (intra-unit cell $C_4$ rotational symmetry breaking) occurs in the cuprates (*10*), with experimental evidence for electronic nematicity coming from bulk transport (*11, 12, 13*) and scanning tunnelling microscopy (STM) measurements (*2, 14).* Despite these developments, the role of electronic nematicity, such as whether it is common to the cuprates and how it relates to CDW order or the crystal structure of different cuprate families, has not yet been established.

Relevant in this context are the stripe-ordered $La_{2-x}M_{x}CuO_4$ (M = Sr, Ba, Eu, Nd) cuprates, where it is known that the low-temperature orthorhombic (LTO) phase to low-temperature tetragonal (LTT) structural phase transition (*15, 16, 17, 18*) plays an important role in stabilizing spin and charge stripe order (*1, 18*). The LTT phase is understood in terms of rotations of the $CuO_6$ octahedra about axes parallel to the Cu-O bonds, with the rotation axis of the octahedra alternating between the $a$ and $b$ axes (*19*) of neighboring planes (Fig. 1)(*15, 16, 17, 18*). This induces $C_4$ symmetry-breaking of the average (or intra-unit-cell) electronic structure within an individual $CuO_2$ plane (referred to here as electronic nematicity) and stabilizes stripes whose orientation rotates by 90° between neighboring planes (*1*). The LTT distortions are understood to be a structural phenomenon, occurring in order to alleviate a lattice mismatch between the preferred Cu-O bond-length of the $CuO_2$ planes and the (La,M)-O bond-length in the $(La,M)_2O_2$ layer (*16*). It is unclear, however, whether there are additional contributions to the electronic nematicity of the $CuO_2$ planes distinct from that engendered by a purely structural distortion and possibly generic to underdoped cuprates.

To address this question, we use resonant soft x-ray scattering (RSXS) at different photon energies to probe the relationship between electronic nematicity of the $CuO_2$ planes, structural distortions, and CDW order. As shown in (*20*), with the x-ray photon energy tuned to the O $K$ edge (at an energy that is sensitive to apical oxygen), the (001) Bragg peak can be used to probe the LTO to LTT phase transition. The (001) peak is forbidden in conventional diffraction but is allowed when the photon energy is tuned to an x-ray resonance (*21, 22*). This occurs because on resonance the scattering cross-section develops sensitivity to the orbital symmetry of atoms (*21, 22*), which differs for atoms in neighboring planes in the LTT phase.

Here, we exploit this orbital and element selective sensitivity and measure the (001) Bragg peak at photon energies corresponding to different elements/orbitals (at the Cu $L$, La $M$, Eu $M$ and O $K$ edges). This enables one to differentiate the $CuO_2$ planes from the $(La,M)_2O_2$ layer and isolate the orbital symmetry of the Cu $3d$ states that are most relevant to the low energy physics of the cuprates. We show that at the Cu $L$ edge, the (001) peak intensity corresponds to $C_4$ symmetry-breaking of the Cu $3d$ states, which we identify as a measure of the degree of electronic nematicity of the $CuO_2$ planes.

In Fig. 2 we show the (001) peak intensity, $I_{001}$, as a function of temperature at absorption edges corresponding to different elements/orbitals: La $M$, Cu $L$, Eu $M$ and O $K$. At the O $K$ edge, the in-plane and apical oxygen sites are further distinguished by photon energy, with 528 eV corresponding to in-plane oxygen, O(1), and $\sim$532 eV corresponding to apical oxygen, O(2) (*20, 23*) (see Fig. 1). We investigated 3 samples, $La_{1.475}Nd_{0.4}Sr_{0.125}CuO_4$ (LNSCO), $La_{1.875}Ba_{0.125}CuO_4$ (LBCO) and $La_{1.65}Eu_{0.2}Sr_{0.15}CuO_4$ (LESCO) having different LTT and CDW transition temperatures, $T_{\mathrm{LTT}}$ and $T_{\mathrm{CDW}}$ respectively. Our principle observation is that measurements of $I_{001}$ for atoms in the $CuO_2$ planes (Cu and O(1)) have a distinct temperature dependence from atoms in the $(La,M)_2O_2$ layer (La, Eu and O(2)). Specifically, the former displays a more gradual temperature dependence than the latter, which exhibits a more pronounced 1st-order-like phase transition. We ascribe this difference to an additional electronic nematicity in $CuO_2$ planes beyond the structural distortions affecting the $(La,M)_2O_2$ layer.

The intensity of the (001) peak at Cu $L$ edge directly measures the nematicity of the Cu 3$d$ states (24). Specifically, $I_{\mathrm{Cu},100} \propto |\eta|^2$, where $\eta = f_{aa}(0) - f_{aa}(0.5)$ (or $= f_{bb}(0.5) - f_{bb}(0)$ by symmetry), and $f_{aa}(z)$ and $f_{bb}(z)$ are diagonal components of the atomic scattering form factor tensor (*25*) averaged over all Cu sites in neighboring $CuO_2$ planes at $z = 0$ or $z = 0.5$ (Fig 1). At the Cu $L$ edge, which probes the Cu $2p$ to $3d$ transition, $f_{aa}$ and $f_{bb}$ are directly related to the orbital occupation of the Cu $3d$ states projected onto the $a$ or $b$ axes, respectively. As such, $\eta$ corresponds to a difference in the symmetry of the Cu $3d$ states between planes. However, because $f_{aa}(0.5) = f_{bb}(0)$ by symmetry, one can express $\eta$ as $\eta = f_{aa}(0) - f_{bb}(0)$ and it thus follows that the (001) peak at the Cu $L$ edge measures the electronic nematicity of the Cu 3$d$ states within a single $CuO_2$ plane.

The (001) peak intensity at other absorption edges have similar sensitivity to orbital asymmetry. For the $(La,M)_2O_2$ layer, this orbital asymmetry tracks the ionic displacements. The

(001) peak intensity measured at a photon energy primarily sensitive to apical O exhibits the same temperature dependence as the (1 0 12) Bragg peak from conventional x-ray scattering, which is sensitive to ionic positions (*26*). Our measurements of (001) peak at energies corresponding to La or apical O also show good agreement with conventional x-ray (*26, 27, 28*) or neutron scattering (*1*), reinforcing our view of a linear relationship between structural distortions and the $(La,M)_2O_2$ layer (001) peaks measured at these photon energies.

Coupling of the symmetry of the electronic structure to ionic positions (e.g. displacements of in-plane oxygen along the $c$-axis in the LTT phase) is also expected for the $CuO_2$ planes (*29*). However, the difference in the temperature dependence of the (001) peak between the $CuO_2$ planes and the $(La,M)_2O_2$ layer shows that electronic nematicity of the $CuO_2$ planes is not simply a trivial consequence of distortions of the $(La,M)_2O_2$ layer and that there is an additional, possibly intrinsic, susceptibility to nematic order in the $CuO_2$ planes.

We next explore the relationship between intra-unit cell ($Q_x = Q_y = 0$) $C_4$ symmetry breaking probed by the (001) peak and inter-unit-cell $C_4$ symmetry breaking arising from unidirectional CDW order. In Fig. 3 we compare measurements of the temperature dependence of the (001) and CDW maximum peak intensities at the Cu $L$ edge. An important case is LESCO, where CDW order onsets at temperatures well below $T_{\mathrm{LTT}}$ (Fig. 3B). Here, the key observation is that the Cu (001) peak is enhanced below $\sim$65K, coincidently with the onset of CDW order. This suggests that nematicity of the $CuO_2$ planes and translational symmetry breaking of the CDW order are intertwined, having a cooperative relationship where CDW order enhances nematicity and vice versa. In contrast, the structural distortion of the $(La,M)_2O_2$ layer, measured by the (001) peak at the Eu $M$, La $M$ or O $K$ (apical) energies, exhibits no such enhancement at $T_{CDW}$ (Fig. 2B). Whereas the LTT structural distortion of the $(La,M)_2O_2$ layer plays a significant role in stabilizing CDW order and nematicity in the $CuO_2$ planes, it does not appear to be coupled as strongly with the CDW order as the nematicity of the $CuO_2$ planes is.

In LBCO at a doping of $x = 1/8$, $I_{CDW}$ behaves roughly as the fourth power of $I_{Cu,001}$ ($I_{CDW} \propto I^4_{Cu,001}$) (Fig. 3A). However, this power-law relationship is not generic, being seemingly not applicable to LESCO and LNSCO, and may be co-incidental. As discussed below, it may also be that 1/8th doped LBCO represents a special case where the order parameters (and/or the coupling parameters) for CDW order, $CuO_2$ plane nematicity and structural distortion of the $(La,M)_2O_2$ layer are tuned to a common critical point.

In a minimal Landau theory that captures the essential aspects of the relationship between CDW, nematic and structural orders, we consider two types of order parameters, the electronic nematic order parameter $\eta$ and the charge density wave order parameter, $\Psi_\alpha$ ($\alpha = x, y$). $\eta$ breaks the point group symmetry $C_4$ down to $C_2$. $\Psi_\alpha$, the complex amplitude of the charge density wave, also generally breaks translational symmetry. Given this, a suitable Landau free-energy, $F$, is

$$F = a\eta^2 - w\eta\Phi + r\left(|\Psi_x|^2 + |\Psi_y|^2\right) + g(\eta + \varepsilon\Phi)\left(|\Psi_x|^2 - |\Psi_y|^2\right) + u\left(|\Psi_x|^2 + |\Psi_y|^2\right)^2 + v\left(|\Psi_x|^4 + |\Psi_y|^4\right). \quad (1)$$

Here $\Phi$ is the structural $C_4$ symmetry-breaking associated with distortions of the $(La,M)_2O_2$ layer, representing the 3D structural phase transition with the octahedral tilting axis rotated by 90° between neighboring planes in the LTT phase. $\eta$ and $\Psi_\alpha$ are associated with a single $CuO_2$ plane within the 3D unit cell. The parameters $a$, $w$, $g$, $\varepsilon, r$, $u$ and $v$ are functions of temperature. From experimental observation, $\Phi$ acquires a non-zero value through a first order phase transition at $T_{\mathrm{LTT}}$, as described by a supplementary part of the Landau theory (*16*). We assume that $\varepsilon\Phi \ll 1$ since the "direct" coupling of the electronic $\eta$ order parameter to $\left(|\Psi_x|^2 - |\Psi_y|^2\right)$ can be expected to be much stronger than the atomic/structural coupling between $\Phi$ and the latter CDW terms. As such, we neglect the $\varepsilon\Phi$ term. Also neglected is interlayer coupling of the electronic nematicity beyond that imposed by $\Phi$. Our measurements support this omission being consistent with the LTT distortion inducing long-range ordering of nematic along the *c*-axis. Although limited by the *c*-axis penetration depth of the x-rays, measurements provide no evidence for a temperature dependent *c*-axis correlation length to the (001) peak, which would indicate a significant role of additional interlayer coupling (*24*).

Within the context of this model, the observed differences amongst LBCO, LESCO and LNSCO involve a material and doping dependence of $\Phi$ and also possibly of $w$, the coupling strength between the LTT structural distortion and electronic nematicity. In contrast, we may conjecture that $r$ and $a$ (the CDW and electronic nematic order inverse susceptibilities), are similar for different cuprate materials at the same hole doping. We identify 3 distinct cases in the theory and their possible correspondence to materials. Case 1: $\left(r - \frac{gw\Phi}{2a}\right) < 0$ at $T = T_{\mathrm{LTT}}$. Here, the 1$^{\mathrm{st}}$ order jump in $\Phi$ at $T_{\mathrm{LTT}}$ is sufficient to induce 1$^{\mathrm{st}}$order jumps in $\eta$ and $\Psi_x$. This case may correspond to LBCO with lower doping $x$ = 0.095 or 0.110 (*27)*,and possibly LNSCO presented here (Fig. 3C). Case 2: $\left(r - \frac{gw\Phi}{2a}\right) = 0$ at $T = T_{\mathrm{LTT}}$. For $T < T_{\mathrm{LTT}}$, this coefficient ultimately becomes negative. Here, the 1$^{\mathrm{st}}$ order jump in $\Phi$ moves the system to the critical point of the CDW order. This special case requires fine tuning of parameters but may in fact correspond to LBCO at $\sim 1/8$ doping (Fig. 3A). Whether such tuning plays a role in the enhancement of CDW order in 1/8 doped LBCO, where the longest range CDW order of any cuprate is observed, or the apparent $I_{\mathrm{CDW}}(T) \propto I_{100,\mathrm{Cu}}^{4}(T)$ relationship is presently unclear. Case 3: $\left(r - \frac{gw\Phi}{2a}\right) > 0$ at $T = T_{\mathrm{LTT}}$. For $T = T_{\mathrm{CDW}} < T_{\mathrm{LTT}}$, this coefficient becomes negative. Here the CDW ordering temperature is well separated from the LTT phase transition, as in LESCO (Fig. 3B) or LBCO with $x = 0.15$ (*27*). Additional work is required to further explore this Landau theory and the relationship of its parameters to material properties (structure, doping and disorder) and other cuprates.

Further context is obtained by contrasting these results to prior reports of nematicity in $YBa_2Cu_3O_{6+x}$ (YBCO), $Bi_2Sr_2CaCu_2O_{8+\delta}$ (Bi-2212) and $Na_xCa_{2-x}CuO_2Cl_2$ (NCCOC). Transport measurements of YBCO have shown in-plane anisotropy that is not solely due to crystalline anisotropy (*11, 12, 13*) and x-ray diffraction measurements have shown CDW order to have unidirectional character (*30, 31, 32, 33*). However, the relationship between the orthorhombic structure of YBCO, unidirectional CDW order and intra-unit cell electronic nematicity is not fully resolved. STM measurements on Bi-2212 and NCCOC arguably show more direct evidence for intra-unit cell electronic nematicity distinct from structure (2, *14, 34*). Due to $C_4$ structural symmetry in these materials, nematicity occurs in Ising-like nanoscale domains. The intra-unit cell electronic nematicity identified here in stripe-ordered La-based cuprates likely represents a long-range ordered analog. This suggests that, like CDW order (*1, 2, 3, 4, 5, 6, 7, 8, 9*), intra-unit cell electronic nematicity may be a generic feature of underdoped cuprates.

• The authors acknowledge insightful discussions with J. Tranquada, S. A. Kivelson, L. Taillefer, I. Fischer, K. M. Shen and G. A. Sawatzky. This work was supported by the Canada Foundation for Innovation (CFI), the Canadian Institute for Advanced Research (CIFAR), the Natural Sciences and Engineering Research Council of Canada (NSERC) and the Canada Research Chair Program (M.J.P.G.). Research described in this paper was performed at the Canadian Light Source, which is funded by the CFI, the NSERC, the National Research Council Canada, the Canadian Institutes of Health Research, the Government of Saskatchewan, Western Economic Diversification Canada, and the University of Saskatchewan. The work at Brookhaven National

Labs was supported by the Office of Basic Energy Sciences, Division of Materials Science and Engineering, U.S. Department of Energy, under Contract No. DE-AC02-98CH10886. M.J.P.G. acknowledges support from the Canada Council for the Arts and the Perimeter Institute (PI) for Theoretical Physics. Research at PI is supported by the Government of Canada through Industry Canada and by the Province of Ontario through the Ministry of Economic Development and Innovation. J.G. and M.Z. were supported by the German Science Foundation (DFG) through the Emmy-Noether (GE1647/2-1) and the D-A-CH program (GE 1647/3-1).

Supplementary Materials
Materials and Methods
Supplementary Text
Figs. S1 to S4
References (*35-39*)

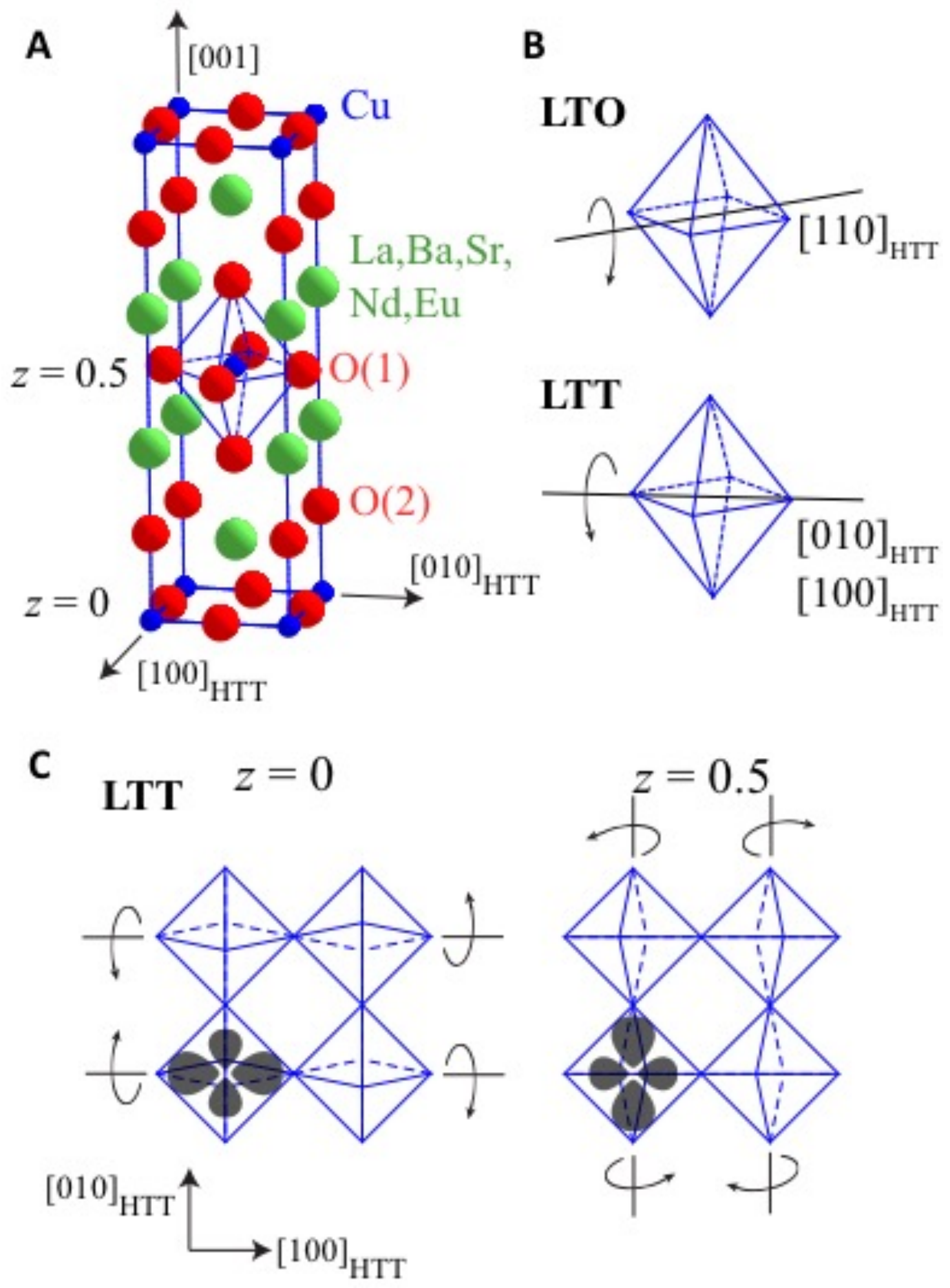

Figure 1: **Structure of La-based cuprates.** **(A)** Unit cell of $(La,M)_2CuO_4$ in the high-temperature tetragonal phase (HTT). O(1) and O(2) are in-plane and apical oxygen sites, respectively. **(B)** Octahedral distortions in the low temperature orthorhombic (LTO) and low temperature tetragonal (LTT) phases. **(C)** $CuO_2$ planes showing the octahedral tilt pattern in neighboring layers ($z$ = 0 and 0.5) in the LTT phase. The structural $C_4$ symmetry breaking and electronic nematicity alternates between neighboring planes, $z$ = 0 and $z$ = 0.5.

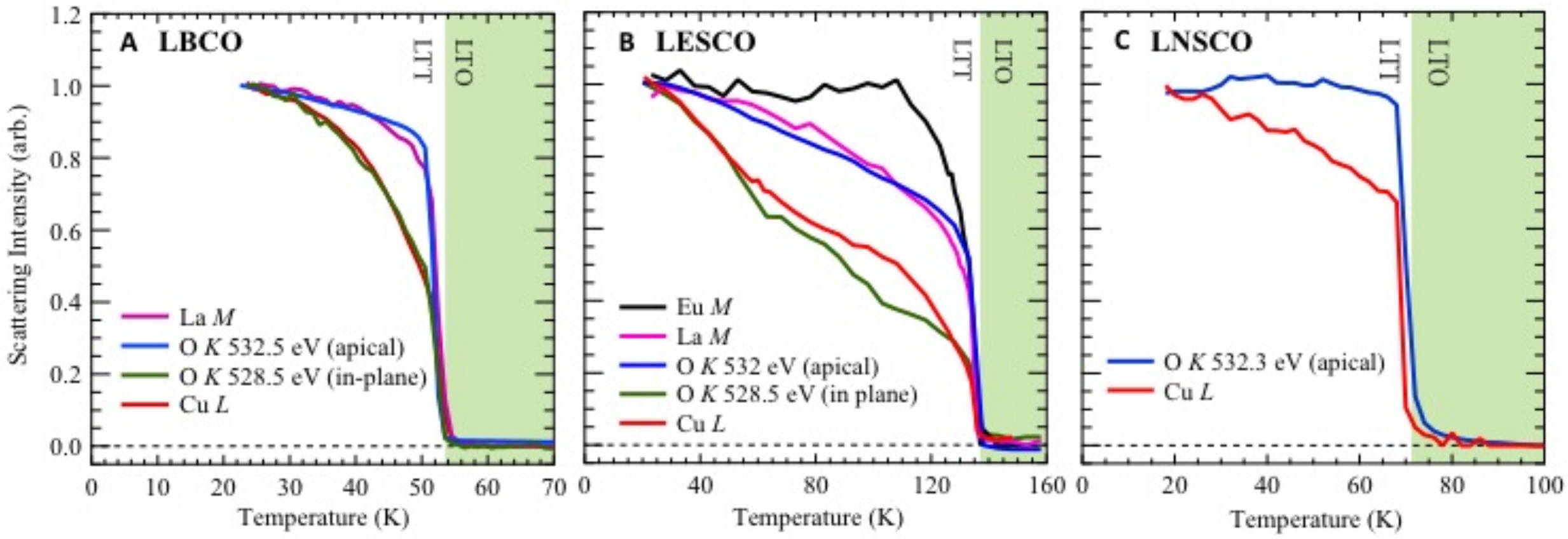

Figure 2: **Temperature dependence of the (001) Bragg peak intensity**. The intensities are normalized by the corresponding low temperature values, $I_{001}(T)/I_{001}(\sim 20\text{K})$, with photon energy tuned to the La $M$, Eu $M$, O $K$ and Cu $L$ edges for **(A)** LBCO, **(B)** LESCO and **(C)** LNSCO. In all cases the (001) peak has a more gradual temperature dependence for Cu and O in the $CuO_2$ planes than for atoms in the $(La,M)_2O_2$ layer.

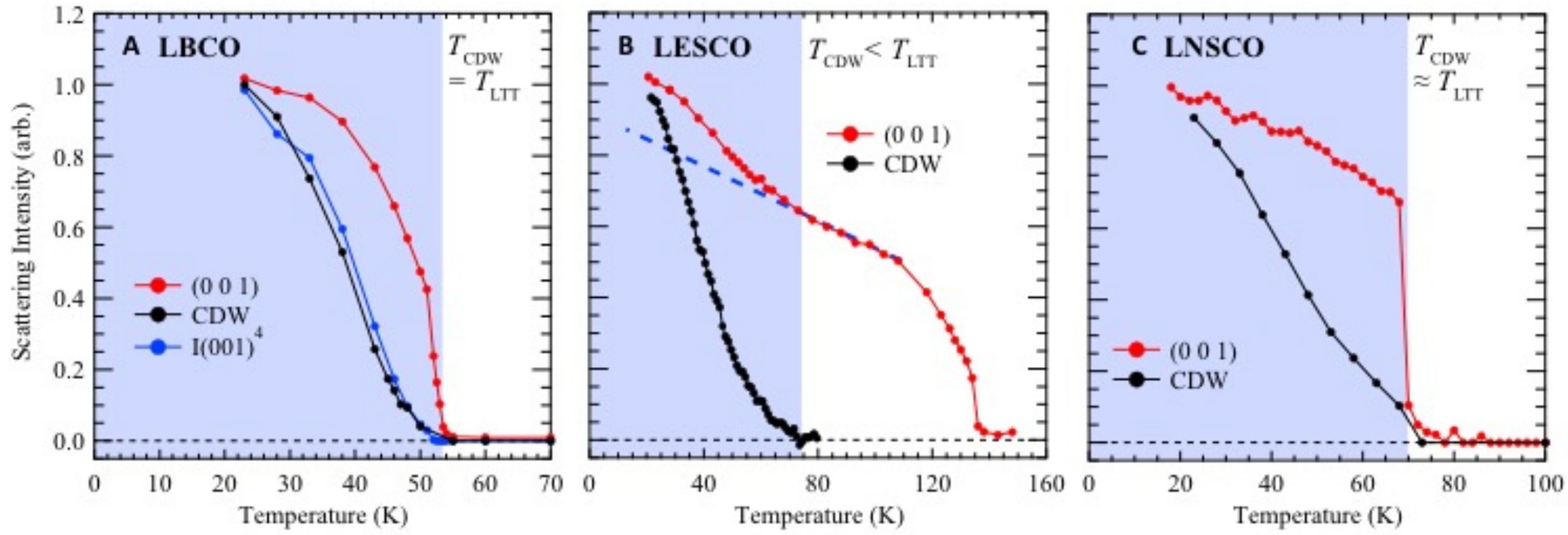

Figure 3: **Comparison between nematicity and CDW order**. Shown is the temperature dependence of the (001) and CDW Bragg peak maximum intensities (normalized by the low temperature value) at the Cu $L$ edge in **(A)** LBCO, **(B)** LESCO and **(C)** LNSCO. In LESCO, the (001) intensity is enhanced below $T_{\mathrm{CDW}}$ indicating a co-operative coupling between nematic and CDW orders. In LBCO, the CDW peak intensity is in good agreement with $I_{\mathrm{Cu},001}^{4}$. CDW peaks were measured at ***Q*** = ($\delta_H$ 0 1.5), where $\delta_H$= -0.238, 0.264 and 0.236 for LBCO, LESCO and LNSCO respectively.

# Supplementary Materials for

## Nematicity in stripe ordered cuprates probed via resonant x-ray scattering

A. J. Achkar, M. Zwiebler, Christopher McMahon, F. He, R. Sutarto, Isaiah Djianto, Zhihao Hao, M. J. P. Gingras, M. Hücker, G. D. Gu, A. Revcolevschi, H. Zhang, Y.-J. Kim, J. Geck, and D. G. Hawthorn correspondence to: dhawthor@uwaterloo.ca.

**This PDF file includes:**

## Materials and Methods

Resonant soft x-ray scattering measurements were performed at the REIXS beamline at the Canadian Light Source (*35*). Samples of $La_{1.475}Nd_{0.4}Sr_{0.125}CuO_4$, $La_{1.875}Ba_{0.125}CuO_4$ and $La_{1.645}Eu_{0.2}Sr_{0.155}CuO_4$ were grown using travelling solvent floating zone and the samples were cleaved out-of-vacuum to expose a fresh surface. The crystallographic orientation of the samples was verified in the diffractometer using the (0 0 2) and (±1 0 3) structural Bragg peaks measured with photon energy >2 keV. Reciprocal lattice units (r.l.u.) were defined using the lattice constants $a = b = 3.787$ Å (3.79 Å) and $c$ =13.24 Å (13.14 Å) for LBCO and LNSCO (LESCO), where the $a$ and $b$ axes are referenced to the high temperature tetragonal unit cell. The CDW and (001) peak measurements were performed with an incident photon energy of 931.3 eV (Cu $L_3$), 1116 eV (Eu $M_5$), 826 eV (La $M_5$ preedge), ~532 eV and ~528.5 eV (O $K$). All temperature-dependent measurements of the CDW and (001) peaks were obtained with σ polarized light with photon polarization along the $b$ axis of the high temperature tetragonal phase. Contributions from the (002) Bragg peak, measured due to $2^{nd}$ order light, were subtracted from (001) peak intensities presented. CDW peak maximum were measured at Q=(δH 0 1.5), where δH = -0.238, 0.264 and 0.236 for LBCO, LESCO and LNSCO respectively. Note the maximum peak intensity is shown in Fig. 3, as opposed to integrated intensity. Additional details of the peak intensities and widths as a function of temperature and photon energy are given in the Supplementary Text below.

## Supplementary Text

Energy dependence of the (001) peak at the Cu $L$ edge

The energy dependence of the (0 0 1) peak was measured at the Cu $L$ edge and is shown in Fig. S1. The scattering intensity is peaked near the Cu $L_3$ and $L_2$ absorption edges (Fig. S1B). The peak width also varies with photon energy through the absorption edges, with an energy dependence similar to the x-ray absorption (Fig. S1C). Moreover, as shown in Fig. S1D, the scattering wavevector of the peak maximum, $\boldsymbol{Q}_{max}$, expressed here in reciprocal lattice units $L_{max} = Q_{max}/(2\pi/c)$, is also energy dependent and deviates from the expected $\boldsymbol{Q} = (0\ 0\ 2\pi/c)$. This occurs due to photon energy dependent refraction of the incident and scattered x-rays at the surface of the sample, leading to a difference between the measured scattering wavevector, $\boldsymbol{Q}$, relative to the momentum transfer within the crystal, $\boldsymbol{Q}'$, that enters into the Laue condition for x-ray diffraction. Similar strong energy dependence to $\boldsymbol{Q}_{max}$, the peak width and the peak intensity were observed at other absorption edges.

These dependencies can be understood within a strong absorption limit of dynamical diffraction theory (*36*) applied to scans along the (0 0 $L$) direction. Here the scattering intensity has a Lorentzian lineshape and is given by

$$\frac{I(\boldsymbol{Q}, E, \epsilon_i, \epsilon_f)}{I_0} = \frac{B^2\left|S(\boldsymbol{Q}', E, \epsilon_i, \epsilon_f)\right|^2}{\left(\pi L - B\mathrm{Re}\left[S(\boldsymbol{0}, E, \epsilon_i, \epsilon_f)\right]\right)^2 - B^2\mathrm{Im}\left[S(\boldsymbol{0}, E, \epsilon_i, \epsilon_f)\right]^2} \qquad \text{(S1)}$$

where $L$ is in reciprocal lattice units ($2\pi/c$), $B = \frac{2c^2r_0}{v_c}$, $c$ is the $c$-axis lattice constant (13.24 Å), is the volume of the unit cell (189.9 Å$^3$), $r_0$ is the Thomson scattering length ($2.82 \times 10^{-5}$ Å), $S(\boldsymbol{Q}', E, \boldsymbol{\epsilon}_i, \boldsymbol{\epsilon}_f) = \sum_j f_j(E, \boldsymbol{\epsilon}_i, \boldsymbol{\epsilon}_f) e^{i\boldsymbol{Q}'\cdot \boldsymbol{r}_j}$ is the structure factor ($\boldsymbol{Q}' = (0, 0, 2\pi/c)$ for the (001) peak), $E$ is the photon energy and $\boldsymbol{\epsilon}_i(\boldsymbol{\epsilon}_f)$ is the incident (scattered) photon polarization. $S(0, E, \boldsymbol{\epsilon}_i, \boldsymbol{\epsilon}_f)$ is the $\boldsymbol{Q}' = 0$ structure factor and is related to the index of refraction. Note, the x-ray momentum transfer within the sample, $\boldsymbol{Q}'$, is related to the measured scattering wavevector $\boldsymbol{Q}$ by $\boldsymbol{Q}' = \boldsymbol{Q} - B\mathrm{Re}[S(0, E, \boldsymbol{\epsilon}_i, \boldsymbol{\epsilon}_f)]/\pi$. $\mathrm{Im}[S(\boldsymbol{0}, E, \boldsymbol{\epsilon}_i, \boldsymbol{\epsilon}_f)]$, and consequently full-width half maximum (FWHM) of the peak, is proportional to the x-ray absorption coefficient, $\mu(E, \boldsymbol{\epsilon}_i, \boldsymbol{\epsilon}_f)$.

As such, the peak width in $L$ is not related to the correlation length of the LTT order, as it does in kinematical diffraction, or to the magnitude of the structure factor, as it does in dynamical diffraction from a perfect crystal without absorption. Also, in this strong absorption limit of the dynamical diffraction, the peak intensity is proportional to $|S(\boldsymbol{Q})|^2/\mu^2$ (as opposed to kinematical diffraction where $I(\boldsymbol{Q})/I_0 \propto |S(\boldsymbol{Q})|^2/\mu$).
Fig. S1C and D show quantitative comparisons of Eqn. S1 to the measured peak FWHM and position, $Q_{\mathrm{max}}$, are shown. For these calculations, $F(0) = 3.75f_{\mathrm{La}} + 0.25f_{\mathrm{Ba}} + 2f_{\mathrm{Cu}} + 8f_{\mathrm{O}}$, where $f_{\mathrm{La}}$, $f_{\mathrm{Ba}}$ and $f_{\mathrm{O}}$ are from Ref. (*36*) and $f_{\mathrm{Cu}}(E)$ is taken from Fig. 2c) of Ref. (*37*). The measured and calculated energy dependences agree within ~ 20%.

Temperature dependence

The temperature dependence of the (0 0 1) peak and (-0.236 0 1.5) CDW peak at the Cu $L$ edge for LBCO are shown in Fig. S2. Consistent with the (0 0 1) peak width being determined by $\mu(E, \boldsymbol{\epsilon}_i, \boldsymbol{\epsilon}_f)$ and not the correlation length of the nematicity, the (0 0 1) peak width is temperature independent (Fig. S2C). Similar temperature independent peak widths are seen at other absorption edges. We note that a small peak remains above 60 K. This peak is due to the (0 0 2) structural Bragg peak, seen here because of a small degree of 2nd order (1862.8 eV) light in the incident beam. In contrast to the (0 0 1) peak, the width of the CDW peak is representative of short range order and exhibits a broadening with increasing temperature (*28*).

Azimuthal angle dependence

At the Cu $L$ edge, the (001) peak intensity can be directly associated with electronic nematicity of the Cu 3*d* states as opposed to lattice displacements or charge modulations along the c axis. It is noteworthy that the Cu position within the unit cell is not changed through the LTO to LTT phase transition and therefore lattice displacements do not enter directly into the structure factor for the (001) peak, which has as simple form:

$$S(001)_{\mathrm{Cu}} = \boldsymbol{\epsilon}_f^* \left(F_{\mathrm{av}}(z=0) + e^{i\pi} F_{\mathrm{av}}(z=0.5)\right) \boldsymbol{\epsilon}_i, \qquad \text{(S2)}$$

where

$$F_{\mathrm{av}}(z) = \begin{bmatrix} f_{aa,\mathrm{av}}(z) & 0 & 0 \\ 0 & f_{bb,\mathrm{av}}(z) & 0 \\ 0 & 0 & f_{cc,\mathrm{av}} \end{bmatrix}, \qquad \text{(S3)}$$

is a tensor representation of the average atomic scattering form factor for Cu in neighbouring planes $z = 0$ and $z = 0.5$. Off resonance, $F_{\mathrm{av}}(z=0)$ and $F_{\mathrm{av}}(z=0.5)$ are identical and the scattering intensity of the (001) peak is zero. However, on resonance they are inequivalent. By symmetry, $f_{cc,\mathrm{av}}(z=0) = f_{cc,\mathrm{av}}(z=0.5)$, $f_{aa,\mathrm{av}}(z=0) = f_{bb,\mathrm{av}}(z=0.5)$ and $f_{bb,\mathrm{av}}(z=0) = f_{aa,\mathrm{av}}(z=0.5)$. Accordingly, the (001) scattering intensity, $I(001) \propto |S(001)|^2$, will be given by:

$$I(001) \propto \left| \boldsymbol{\epsilon}_f^* \begin{bmatrix} \eta & 0 & 0 \\ 0 & -\eta & 0 \\ 0 & 0 & 0 \end{bmatrix} \boldsymbol{\epsilon}_i \right|^2, \qquad \text{(S4)}$$

where $\eta = f_{aa,\mathrm{av}}(z=0) - f_{aa,\mathrm{av}}(z=0.5) = f_{aa,\mathrm{av}}(z) - f_{bb,\mathrm{av}}(z)$ is a measure of the electronic nematicity within a given plane.

The validity of Eqn. S4 can be verified experimentally by measuring the dependence of the scattering intensity as the crystalline axes of the sample are rotated azimuthally relative the incident photon polarization, which can be oriented parallel ($\pi$) or perpendicular ($\sigma$) to the scattering plane as depicted in Fig. S3B. The azimuthal dependence of $I_\pi/I_\sigma$ agrees well with the scattering intensity calculated from Eqn. S4, with no free parameters, providing strong confirmation that the (001) peak at Cu is measuring electronic nematicity and has a form given by Eqn. S4. Note, here the calculated scattering intensities are corrected for the azimuthal angle dependence of the x-ray absorption by dividing by $\mu(931.3\text{ eV}, \boldsymbol{\epsilon}_i, \boldsymbol{\epsilon}_f)^2$ (*32*).

In contrast, other types of order would provide a very different azimuthal dependence. For illustrative purposes, we also show in Fig. S3A the azimuthal dependence that would correspond to modulations between planes of the average charge or lattice position (here the unoccupied orbital occupation of the Cu $3d_{3z^2-r^2}$ states is taken as 10% of the Cu $3d_{x^2-y^2}$ states) or to antiferromagnetic ordering of the average *c*-axis component of the Cu magnetic moment along the *c*-axis. These types of order have scattering tensors given by:

$$I(001)_{\text{charge/lattice}} \propto \left| \boldsymbol{\epsilon}_f^* \begin{bmatrix} 1 & 0 & 0 \\ 0 & 1 & 0 \\ 0 & 0 & 0.1 \end{bmatrix} \boldsymbol{\epsilon}_i \right|^2, I(001)_{m_c} \propto \left| \boldsymbol{\epsilon}_f^* \begin{bmatrix} 0 & 1 & 0 \\ -1 & 0 & 0 \\ 0 & 0 & 0 \end{bmatrix} \boldsymbol{\epsilon}_i \right|^2. \qquad \text{(S5)}$$

<u>X-ray absorption: O *K* edge</u>

The x-ray absorption (XAS) measured at the O *K* edge is shown in Fig. S4 for LNSCO, LBCO and LESCO using total fluorescence yield. The XAS is qualitatively very consistent between the 3 samples. At ~528.5 eV, the electronic states are dominated by the O 2*p* states of the in-plane oxygen (*39*). In contrast, the peak at ~532 eV, is characterized by apical oxygen, O(2), hybridized with La, Ba, Sr, Eu and Nd (*20, 23*).

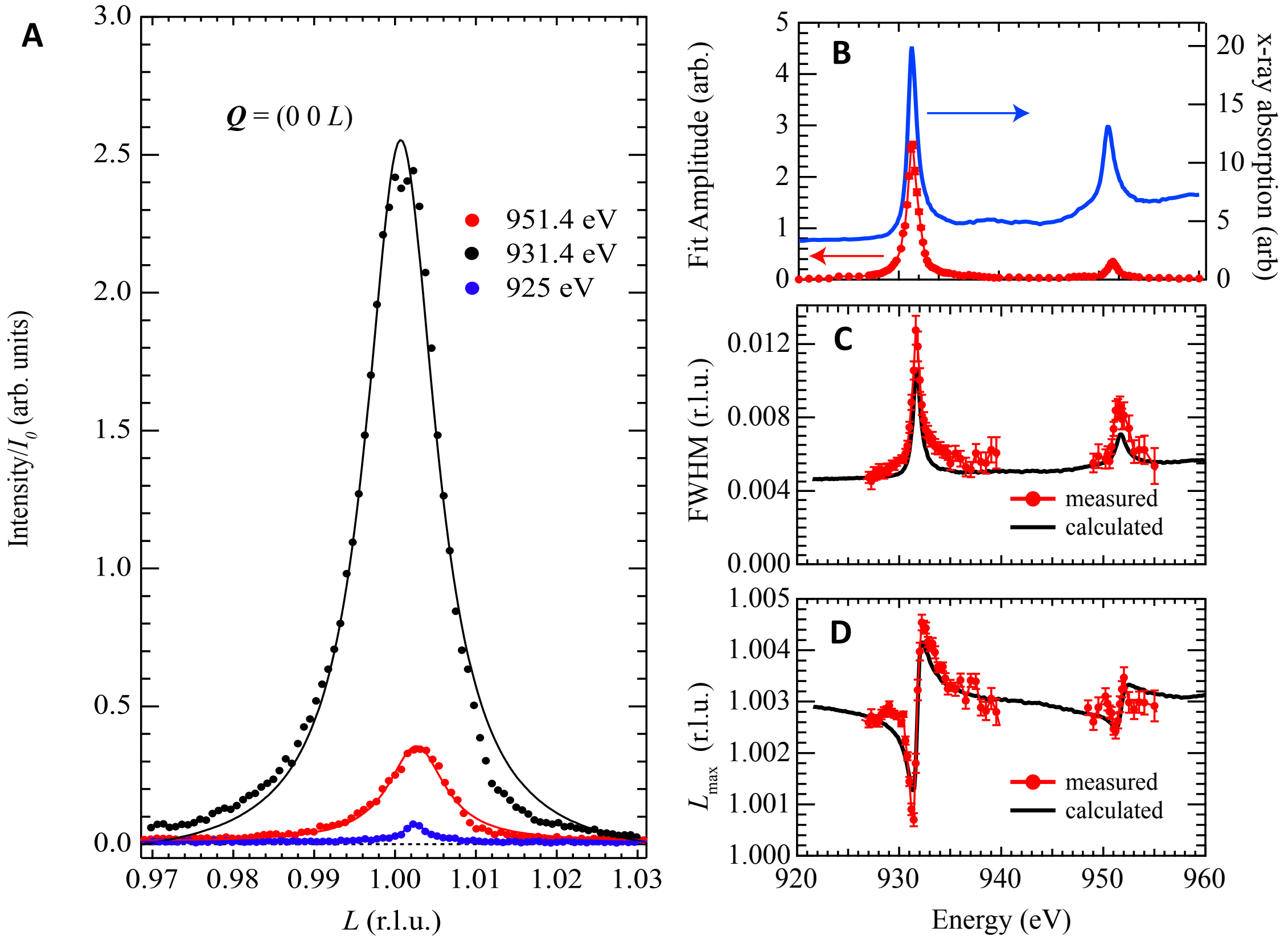

**Fig. S1. The energy dependence of the (001) reflection in LBCO measured at the Cu *L* edge.** (A) The measured peak intensity (circles) at select photon energies as a function of scattering wavevector $Q$ scanned along the (0 0 $L$) direction, with $Q$ expressed in reciprocal lattice units (r.l.u.), $2\pi/c$. Solid lines are Lorentzian fits to the data. (B) The fitted peak amplitude (red) and x-ray absorption (blue). X-ray absorption is measured with total fluorescence yield and photon polarization $\boldsymbol{\epsilon_i}\ //\ \boldsymbol{a}$. (C) The measured (red) and calculated (black) FWHM of the peak. (D) The measured (red) and calculated (black) peak maximum scattering vector, $Q_{\max}$, in reciprocal lattice units.

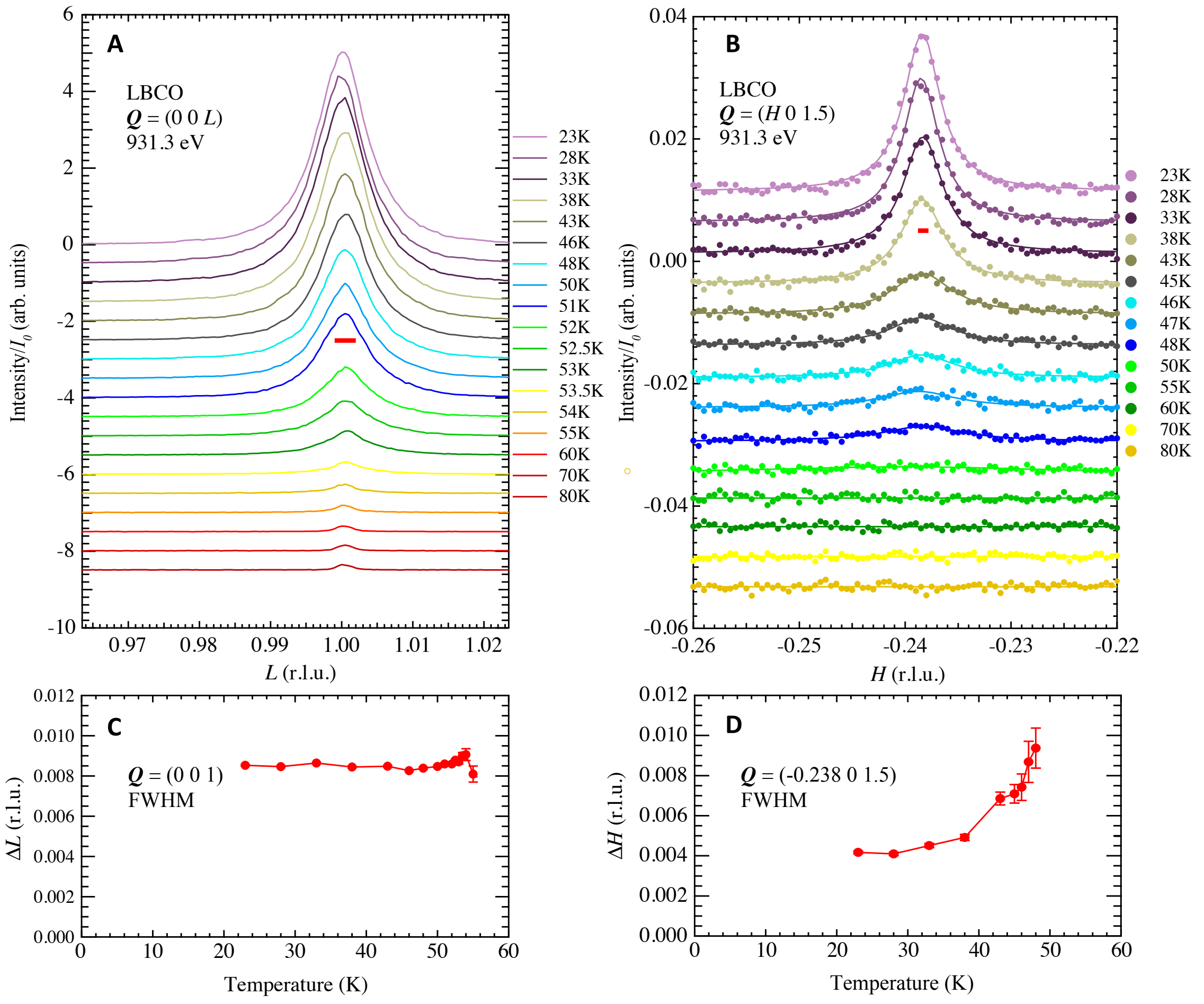

**Fig. S2 Temperature dependence of the (001) and (-0.236 0 1.5) CDW peaks in LBCO measured at the Cu L edge (931 eV).**

**(A)** Intensity as function of *L* at various temperatures for the (001) peak. The peak intensity decreases with increasing temperature, but the width remains constant. Above 60 K, a small peak remains due to the (002) Bragg peak, observed here due to second order light ($E_i$ = 1862 eV). **(B)** Intensity as a function of *H* at various temperatures for the (-0.236 0 1.5) CDW peak. In **(A)** and **(B)** the red horizontal bars indicate the experimental resolution. **(C)** and **(D)** The peak width (FWHM) as a function of temperature for the (0 0 1) and (-0.236 0 1.5) peaks.

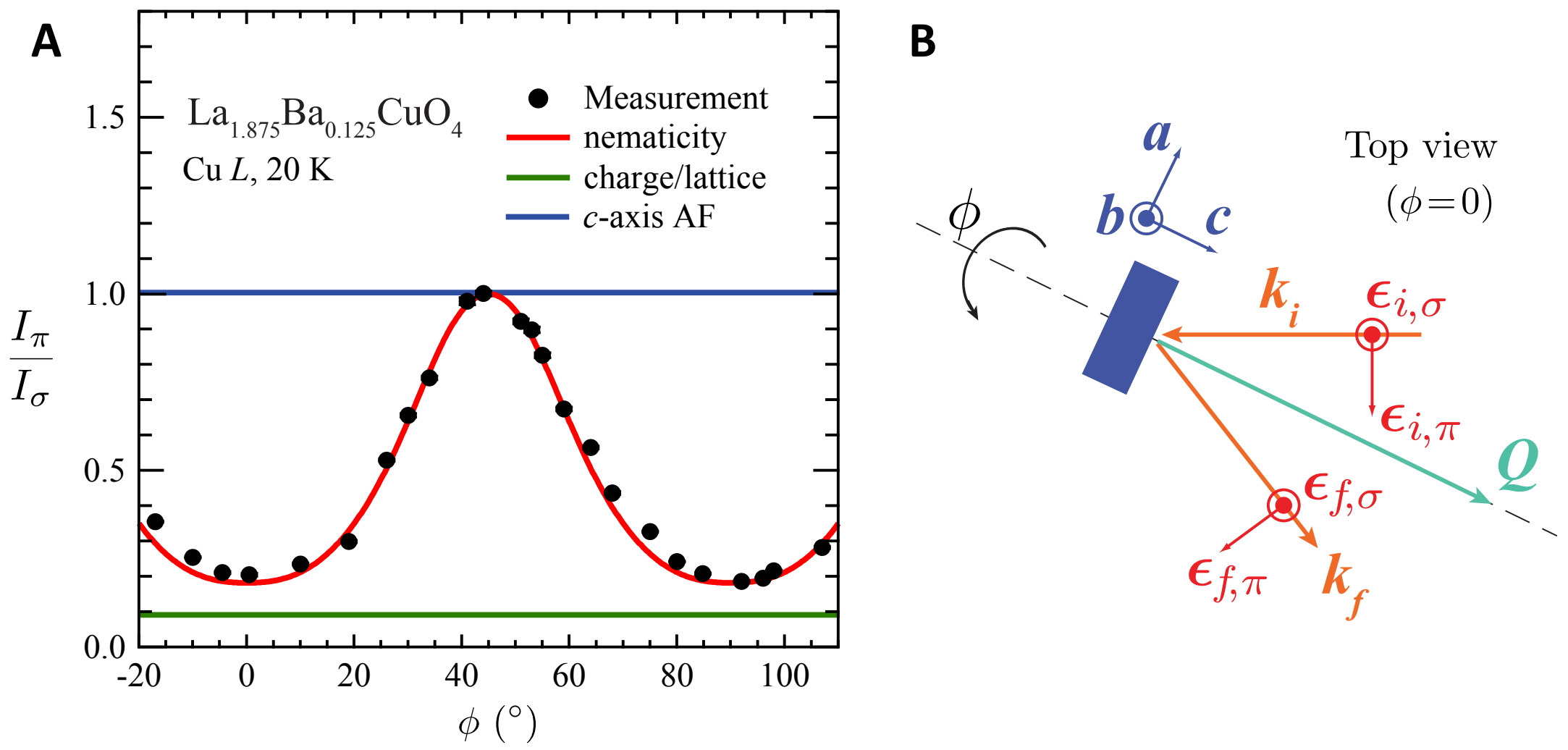

**Fig. S3**

**(A)** The azimuthal ($\phi$) dependence of $I_\pi/I_\sigma$, the ratio of the (001) Bragg peak scattering intensity with π and σ incident polarization, measured at the Cu *L* edge (931.3 eV). The red, green and blue lines are calculation of the scattering intensity to the forms given in Eqn.'s S4 and S5 divided by $\mu(931.3\text{ eV}, \boldsymbol{\epsilon}_i, \boldsymbol{\epsilon}_f)^2$. **(B)** A top view of the experimental measurement geometry. With $\phi = 0$ and 90°, the Cu-O bond is parallel to the scattering plane.

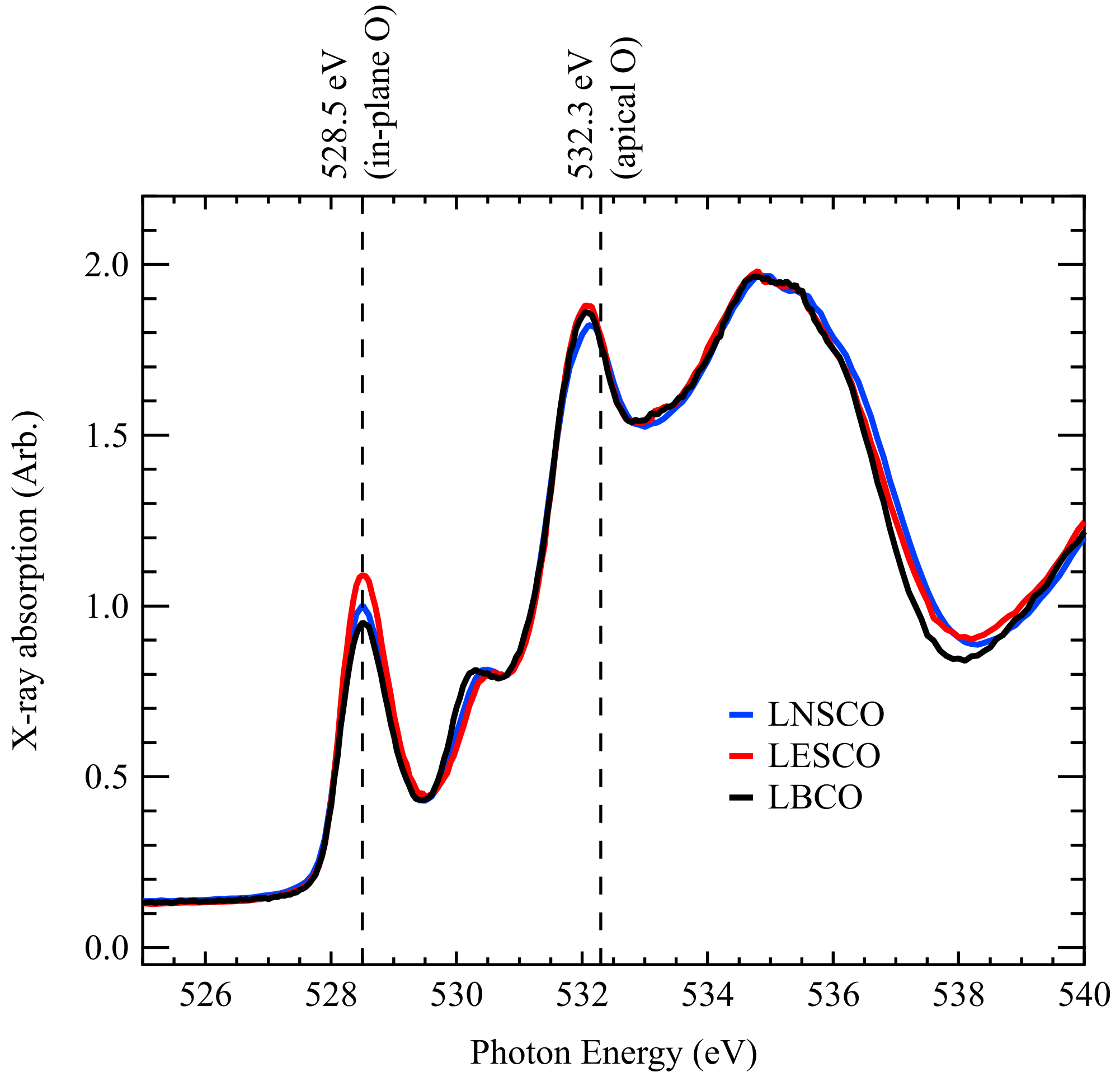

**Fig. S4**

The x-ray absorption at the O *K* edge in LBCO, LNSCO and LESCO measured using total fluorescence yield. At 528.5 eV, the XAS is dominated by in-plane, O(1), states (*39*) whereas at ~532 eV, the XAS is influenced by apical oxygen, O(2), states hybridized with (La,M) states (*20, 23*). The measurements on the three samples have been normalized above and below the absorption edge.